\documentclass[runningheads]{llncs}
\usepackage[T1]{fontenc}
\usepackage{graphicx}
\usepackage{comment}

\usepackage{tikz}
\usetikzlibrary{positioning}
\usetikzlibrary{shapes.geometric, arrows.meta, positioning}
\tikzstyle{stepbox} = [rectangle, rounded corners, minimum width=2.5cm, minimum height=0.75cm, text centered, font=\normalsize]
\tikzstyle{arrow} = [thick,->,>=stealth]

\usepackage{caption} 
\usepackage{hyperref}
\usepackage{float}
\begin{document}
\title{Eliciting User Requirements for AI-Enhanced Learning Environments using a Participatory Approach}
%
%
\author{Bibeg Limbu\inst{1}\orcidID{0000-0002-1269-6864} \and
Irene Angelica-Chounta\inst{1}\orcidID{0001-9159-0664} \and Vilma Sukacke\inst{2}\orcidID{0000-0002-8990-019X} \and Andromachi Filippidi\inst{3}\orcidID{0009-0000-1816-7907} \and Chara Spyropoulou\inst{4} \and 
Marianna Anagnostopoulou\inst{4} \and Eleftheria Tsourlidaki \inst{5}\orcidID{0000-0001-9568-5935} \and Nikos Karacapilidis\inst{3}\orcidID{0000-0002-6581-6831}}
\authorrunning{B. Limbu et al.}
%
\institute{University of Duisburg-Essen, Germany \and Kaunas University of Technology, Lithuania \and University of Patras, Greece \and IASIS, Greece \and Uni Systems, Greece}
\maketitle              
%
\begin{abstract} 
This paper explores the needs \& expectations of educational stakeholders for AI (Artificial Intelligence)-enhanced learning environments. Data was collected following two-phased participatory workshops. The first workshop outlined stakeholders' profiles in terms of technical and pedagogical characteristics. The qualitative data collected was analysed using deductive thematic analysis with Activity Theory, explicating the user needs. The second workshop articulated expectations related to the integration of AI in education. Inductive thematic analysis of the second workshop led to the elicitation of users' expectations. We cross-examined the needs \& expectations, identifying contradictions, to generate user requirements for emerging technologies. The paper provides suggestions for future design initiatives that incorporate AI in learning environments.

\end{abstract}
\section{Introduction}


The integration of AI in Education (AIED) should follow human-centered design principles, thus prioritizing human needs and capabilities, rather than replacing human roles or undermining human agency \cite{holmes2022artificial,chounta_exploring_2022}. This necessitates interdisciplinary collaboration \cite{dickler2022interdisciplinary}. To promote human-centered design of AI-enhanced learning environments, the involvement of educational stakeholders, such as teachers and learners, in the design process of these environments is necessary \cite{topali2025designing}. At the same time, the integration should be guided by established learning and pedagogical theories
\cite{chounta_exploring_2022,Lavidas_2024_determinants}. To that end, participatory design is an approach that involves the active and meaningful participation of stakeholders, including end-users, in the design process. It aims to ensure that the solutions meet the needs and expectations of the end users \cite{kensing1998participatory}. This methodology emphasizes collaboration, co-creation, and shared decision-making between designers and stakeholders.

We explored two distinct learning settings drawn from the broader project context (see Section \ref{secMethdata}): a higher education course for pre-service teacher training, and a course on Leapfrogging Industry 4.0 technologies designed for civic society watchdogs and EU civilian missions. Stakeholders from these settings were involved in two workshops with the aim of eliciting requirements for the design of an AI-enhanced learning platform for 21st-century skills. Qualitative data collected during the participatory workshops were analyzed using thematic analysis. To guide our analysis, we formulated the following questions:\

\textbf{RQ1:} What are the needs of educators and learners for teaching and learning 21st-century skills?\\ In particular, we want to explore a) the current state of 21st-century skills training in practice; b) the tools and pedagogical approaches used, and c) the challenges faced.
    
\textbf{RQ2:} What are the expectations of educators and learners of emerging technologies for the development of 21st-century skills?\\ We aim to specify a) the challenges in designing learning activities, and how emerging technologies can solve them, and b) how educators monitor learner performance, what evaluation methods they use, and what types of AI-powered recommendations would they find valuable?


\section{Related work}

\subsection{Participatory design}\label{secParticipatory}
We organised two participatory workshops to collect data to elicit user requirements. Participatory design is an approach that prioritizes the active involvement of participants, particularly end-users, in the design and development process \cite{muller_participatory_2012}. Participatory design also involves end-users, stakeholders, and other relevant parties in the decision-making processes of a product, service, or system. Participants, as active contributors, offer insights to guide the development process, an element that characterizes this methodology \cite{spinuzzi2005methodology}. By fostering collaboration, co-creation, and shared ownership among stakeholders, participatory design aims to develop solutions that better address real-world challenges and meet the diverse needs of the community. It aims to create inclusive, user-centered, and contextually relevant solutions by actively engaging those affected by or benefiting from the design outcomes. Participatory design recognizes the expertise and lived experiences of the participants, enabling them to contribute their perspectives throughout the design process \cite{ElHelou_Tzagarakis_Gillet_Karacap_Yu_2008}. This approach can lead to greater user satisfaction, increased usability, and more socially responsible outcomes.

\subsection{Activity Theory}\label{secAcT}

\begin{figure}
\centering

\includegraphics[width=0.9\textwidth]{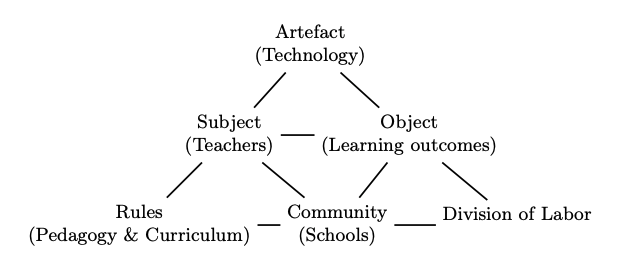}

\caption{Uden \& Ching’s \cite{uden_activity_2024} Ecosystem for AIED design}\label{Fig_ActivityAIED}
\end{figure}
Activity Theory (AcT) emphasises a holistic "ecosystem" perspective in the requirements analysis, instead of the excessive focus on technology put by other models like the TPACK model \cite{uden_activity_2024}. It prioritizes the importance of studying the real-life use of technology as part of the unfolding human interaction with the world over the traditional HCI methods, which focus on formal, abstract models of the user and the system as separate entities within the interaction. AcT facilitates a rich, systemic understanding of users' needs embedded within their actual work, tools, rules, and communities \cite{Engeström_2014}. Using AcT in the analysis of user requirements supports the design of solutions that address the actual underlying problems and constraints, thus improving the relevance and sustainability of the solutions \cite{clemmensen_making_2016}.

AcT conceptualizes \textit{activity systems} as the fundamental unit of analysis for examining complex, real-life human practices. An activity system is a dynamic, historically evolving, goal-directed, and tool-mediated structure of human practice \cite{Engeström_2014}. Following Uden \& Ching's \cite{uden_activity_2024} interpretation of AcT in the context of AIED (see Figure \ref{Fig_ActivityAIED}), AcT presupposes that the relationship between the \textbf{Subject} (Teachers) and the \textbf{Object} (Learning Outcomes) is mediated by the \textbf{Artifacts} (Technology, including AI). This activity system is situated within a \textbf{Community} (Schools), which is governed by a set of \textbf{Rules} (Pedagogy \& Curriculum) and organized through a \textbf{Division of labour}. 

\section{Methodology}\label{secMethdata}

\begin{figure}
\centering
\includegraphics[width=0.8\textwidth]{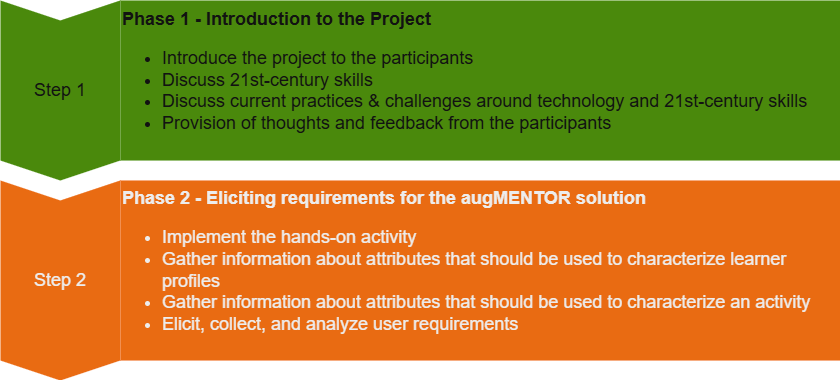}
\caption{Workshops structure per phase of implementation.} \label{fig_Method}
\end{figure}
We adopted a participatory approach in line with the assumptions of AcT. In particular, we invited stakeholders from two pilots of a European research project (augMENTOR\footnote{\url{https://augmentor-project.eu}}) to participate in this study. These stakeholders consist of higher education teachers, teacher trainers, and policymakers with extensive experience in distance learning and the use of technologies in their daily teaching practice. The pilots were organized and led by Kaunas University of Technology (Pilot 1) and the University of Patras (Pilot 2). The structure of the study is presented in Figure \ref{fig_Method}; each pilot conducted one workshop per phase (two in total). The workshops’ structure was the same for all pilot partners, with minor adjustments to meet the needs of different groups of participants. Storyboards were used to scaffold and communicate the workshop structure between the pilots. Each pilot partner was responsible for facilitating their workshops. The participants from the first workshop were encouraged to participate in the second workshop. This is aimed at maximizing user engagement, facilitating a collaborative atmosphere, and ensuring continuity, enabling participants to build upon earlier discussions. Insights were collaboratively documented using an online whiteboard. In Workshop 1, there were 6 participants in Pilot 1, and 6 participants in Pilot 2. In Workshop 2, there were 8 participants in Pilot 1, and 5 participants in Pilot 2. No personal data was collected during the study. Participants confirmed their voluntary participation with signed consent forms, and the study was approved by the project’s Ethics Advisory Board and Data Protection Officers of the pilot partners.

\section{Results}\label{sec2}
We used thematic analysis to synthesize insights from the collected qualitative data. Thematic analysis is a method for identifying, analysing, and reporting patterns (themes) within data \cite{braun_using_2006}. In this paper, we employed Braun and Clarke \cite{braun2021thematic}'s 6-phase process for thematic analysis: 1) Familiarisation with the data, 2) Initial coding, 3)Generating themes, 4) Reviewing themes, 5) Defining \& naming themes, and 6) Producing the report. Deductive thematic analysis \cite{braun2021thematic} guided by Activity Theory (AcT) \cite{Engeström_2014,uden_activity_2024}, was used to analyse qualitative data from workshop 1 (see Section~\ref{secAcT}). Google Sheets was used to systematically organize, sort, and review codes and themes.

\subsection{Elicitation of User Needs (Workshop 1)}
\subsubsection{Subjects} represents the actor or group (in this case, educators) engaging in the activity. Analysis of the data revealed two central themes characterizing participants' orientations and experiences: Beliefs and Attitudes (\textbf{Theme 1}), and Pedagogical Preparedness (\textbf{Theme 2}).

In \textbf{Theme 1}, educators expressed a range of beliefs and their dispositions that shape their engagement with educational technology and 21st-century skills development. A common thread was a \textbf{resistance to change}, with participants describing hesitation or reluctance to deviate from familiar tools or teaching routines, even when new options were available [\textit{"The major problem is to persuade teaching staff of the importance of these skills"}]. This resistance was often also tied to a sense of comfort with existing practices or fear of complexity [\textit{"(Teacher)..keep using their standard tools they feel comfortable with, they do not desire new tools"}]. Additionally, \textbf{distrust of technologies} surfaced, particularly around newer technologies [\textit{"how do we (teachers) judge AI assessment?"}]. 

\textbf{Theme 2} highlights the challenges in educators' readiness to implement 21st-century teaching practices, particularly when it comes to integrating new technologies. One prominent concern was the \textbf{lack of knowledge on 21st-century pedagogies}. Educators felt underprepared to facilitate 21st-century skills development with emerging technologies [\textit{"the majority of our teaching staff is not familiar with the terms and their development"}]. This was compounded by the conflict generated by the \textbf{lack of support} provided by the  Institutions and Organisations (see Section \ref{secCommunity}), with educators often reporting being left to navigate technological adoption on their own [\textit{"teachers need tech support to apply the knowledge and skills gained in training"}]. 

\subsubsection{Objects}\label{secObjects} of an activity represents the underlying purpose or intended outcome that drives behavior. The themes underlying the participants' objectives are: Transfer of Skills (\textbf{Theme 1}), Equity in Learning (\textbf{Theme 2}), and Professional Development (\textbf{Theme 3}).

\textbf{Theme 1} highlights the teachers' objective to ensure that learners can meaningfully \textbf{apply what they’ve learned}. For instance, participants highlighted that educational tools should help reflect on and transfer new knowledge to their professional or everyday practice [\textit{"People get training (to include 21st century skills) but are left alone in how to apply the new knowledge and skills"}]. Correspondingly, there was also recognition of the importance of follow-up support, where tools would continue to guide post-training [\textit{"I need an AI tool to help me follow up on what I've learned and guide me on how to implement that in practice"}].

\textbf{Theme 2} reflects the teacher's intention to \textbf{promote equitable learning}, primarily through personalised instruction and addressing technological literacy barriers. Several participants expressed the desire for AI tools to provide "individual learning experiences" [\textit{"(classes)..are usually homogeneous, it is important to target the training to individuals, which could be done with technology"}]. Concerns about digital literacy levels also surfaced, particularly as the teachers do not have the time to focus on it [\textit{"A certain level of (tech) literacy skills is needed to attend the training, but it is not what we (teachers) focus on in our training"}]. 

\textbf{Theme 3} emphasises the educators' drive for their own \textbf{professional development}. Some educators viewed technology use as a way to stand out [\textit{"As an individual trainer, it is important to stand out and show trainees something new"}]. Others framed it as part of their own professional goals, seeking to align their practice with evolving expectations for digital competence. For example, teachers are expected to be compliant with institutional or legal frameworks, such as GDPR. Educators desire tools that offload tasks, for example, tools that are standards compliant by design [\textit{"It would be fantastic to have a GDPR tool inbuilt in whatever tech we use, there is no solution for this yet"}].

\subsubsection{Tools} represents the \textbf{tools} (technologies in this instance) that subjects use to achieve their objects. Tools directly influence practice, shaping how educators can facilitate 21st-century competencies. We identified: (i) Adoption challenges of emerging technologies (\textbf{Theme 1}), and (ii) Desired features in emerging technologies (\textbf{Theme 2}), which highlights where improvements or interventions with AI-based solutions are needed. 

\textbf{Theme 1} highlights the challenges educators face when adopting new technological tools in their practices. A prominent concern was the \textbf{lack of guidance} for selecting and integrating tools [\textit{"Integration of various applications and devices can cause issues in training..so you (teachers) avoid it"}], leaving educators to self navigate with uncertainty [\textit{"I am constantly figuring out how to use more technology or which ones to choose for a specific purpose}"]. Additionally, the educators and students cannot consistently afford the \textbf{cost} of licenses or subscriptions for tools [\textit{"<LMS> is free, so it was an advantage, good tools become expensive or change their features"}]. 

\textbf{Theme 2} captures the key functionalities that educators desire in tools, particularly those enhanced by AI.
A recurring need was \textbf{personalisation} [\textit{"(Teachers need) tools that could provide one-on-one attention}]. The educators emphasized facilitating learning beyond the classroom, supporting application of knowledge, thereby helping learners bridge the gap between training and real-world practice [\textit{"If the tool could help follow up on training, it would be useful because there is a gap between the course and application of knowledge"}]. Similarly, participants stressed the value of authentic learning support with technologies to better support transfer of knowledge [\textit{"Simulations may provide virtual environments where students can practice, test hypothesis, experiment"}] despite their reservation on its efficacy [\textit{"transferring personal experience to virtual reality is challenging"}]. Finally, participants also expressed interest in \textbf{task offloading}, where tools assist not just with routine or administrative responsibilities, but higher cognitive tasks such as instructional design [\textit{"I want a tool that matches learning objectives with appropriate teaching methods, such as theoretical teaching and practical activities"}]. 

\subsubsection{Rules}\label{secRules} encompass the formal policies and informal norms that regulate the activity. Rules influence how educators adopt, avoid, or adapt tools in their practice. The analysis revealed four interrelated categories of rules: pedagogical (\textbf{Theme 1}), institutional and societal (\textbf{Theme 2}), and functional (\textbf{Theme 3}).

\textbf{Pedagogical norms} highlighted the educator's struggle in navigating ambiguous or underdeveloped pedagogical guidance, specifically regarding the 21st-century skills. Educators were unsure how to incorporate 21st-century skills meaningfully within their teaching [\textit{"A long list of 21st century skills that is constantly updated is not helpful. They need to be organised so that they can be acquired in real-life"}]. While these competencies were valued, the participants also noted the lack of assessment policies focused on 21st-century skills [\textit{"We do not necessarily measure their critical thinking, what we do asses is their (students') needs"}].

\textbf{Institutional policies and societal norms} shape and constrain how educators engage with tools. Educators adhere to the top-down imposition of tools by their institutions [\textit{"we (teachers) use chalkforms for quizzes, also Moodle, and whatever tools are available}]" or the societal norms [\textit{"Around 2010s, I (a teacher) started using prezi presentations, but around 2015 I resorted back to powerpoint presentation because the entire world was using it"}], reflecting their tendencies to align with common accepted practices, regardless of pedagogical fit. Policies such as data protection regulations also shape practice [\textit{"we (teachers) used a lot of Google features, but we are moving to <company name> because of data protection"}]. However, participants described how the development of 21st century skills are oten referenced in policies, indireclty supporting it but only in symbolic terms, leading to a disconnect between institutional rhetoric and actual classroom implementation [\textit{"The university highlights and supports their (21st century skills) development through research programs and staff trainings, however in practice the result does not correspond to these efforts"}]. 


\textbf{Functional norms} describe the pragmatic constraints that act as powerful informal rules. Participants reported using technology purely \textbf{based on its purpose}. For instance, participants described using Zoom because it facilitates conventional lectures. Educators simply viewed technology \textbf{as a medium} [\textit{"we use tools to compensate for the lack of f2f learning"}]. Many also described a fragmented digital environment, where the sheer number of tools and platforms creates confusion, discouraging their integration [\textit{"Integration of various applications and devices can cause issues, so you (teachers) avoid it"}]. Finally, a widely echoed sentiment emerged, "if it’s not broken, don’t fix it," where any deviation from established practice was viewed as unnecessary [\textit{"not getting feedback from students keeps trainers comfortable in using something they are used to"}].

\subsubsection{Community}\label{secCommunity} represents the stakeholders who share the same object (goal) as the subject, and(or) influence(ed) by the activity. We identified three main themes under community: Students (\textbf{Theme 1}), Institutions (\textbf{Theme 2}), and Pedagogy researchers (\textbf{Theme 3}).

Educators reported that students (\textbf{Theme 1}) often indirectly determined the instructional settings through their needs [\textit{"Students' competencies, based on that, content is chosen and developed"}] or logistical constraints (e.g., number of students). A recurring challenge was the homogeneity of the cohort, which made it difficult to tailor instruction to individual students [\textit{"Students groups are usually homogeneous, it is important to target individuals"}]. Educators also emphasized the students' preference for face-to-face learning, citing the importance of human experience and emotional communication. These expectations limit the perceived adequacy of digital tools [\textit{"Students are hungry for f2f learning, technology cannot provide this human experience"}]. 

Institutions (\textbf{Theme 2}) played a significant role in shaping the activity. In many cases, institutional decisions imposed constraints that indirectly but strongly influence how educators plan, deliver, and adapt technologies to their teaching. However, institutions fail to support their directives. As one participant explained, [\textit{"Teachers get training (to include 21st century skills) but are left alone in how to apply the new knowledge"}]. In line with Institutional norms (see Section \ref{secRules}), Institutions impose technological environments without sufficient consultation with educators or pedagogy in mind. 

Pedagogy researchers (\textbf{Theme 3}) influence educational activities, but Participants expressed concern that researchers, while producing valuable theoretical insights, often struggled to convey their significance [\textit{The major problem is to persuade teaching staff of the importance of these skills}]. This gap was more pronounced in the domain of 21st-century skills, where educators called for more pragmatic approaches to train them [\textit{"A long list of 21st century skills that is constantly updated is not helpful in any educational context. It needs to be organised so that they can be acquired in real-life}"].

\subsubsection{Division of Labour} refers to the distribution of tasks, responsibilities, and authority among people involved in the activity. Two themes emerged from the analysis: Gaps and Overloads (\textbf{Theme 1}) and Unclear Responsibilities (\textbf{Theme 2}). 

\textbf{Theme 1} highlights the disproportionate share of responsibility in ensuring the "objects" of the activity are met. Educators often had to choose, integrate, and justify their selection of tools from a vast array of tools with little guidance or contextual support, describing it as demanding [\textit{"it would be useful to have AI tools that give you (teachers) suggestions on what to use to deliver specific learning scenarios"}]. Educators also noted the complexity and tediouness in translating their in-person teaching into digital mediums [\textit{"transferring the personal experience to virtual reality is challenging"}], especially with emotional nuance [\textit{"Emotional aspect when dealing with tech, in comparison to f2f, is not there}]. This gap reflects an underlying tension in the division of labour, where the expectation to digitize pedagogical presence is placed on teachers, but without adequate systemic scaffolding.


While the importance of AI-driven tools in education was widely acknowledged, a critical barrier emerged in the form of the ambiguity in various roles needed for their successful implementation \textbf{Theme 2}. Participants often noted that although technology was being enthusiastically integrated, it was unclear whose role it was to ensure users had the necessary skills to engage effectively. As one participant put it, [\textit{"My teaching involves 21st-century skills \& digital literacy; however, my university does not formally prioritize these skills"}]. Participants were unsure about who carries out the evaluation of integrated tools regarding their pedagogical effectiveness, ethical alignment, or suitability for learning [\textit{"if AI is involved in assessment, how do you (teachers) judge it?"}]. While students were positioned as sources of informal feedback, there was no formal mechanism or designated actors in place to systematically gather and act on feedback [\textit{"not getting feedback from trainees keeps trainers comfortable in using something they are used to"}].

\subsection{Elicitation of Users' Expectations (Workshop 2)}
We employed inductive data-driven thematic analysis to analyse qualitative data from workshop 2 to analyse user expectations. In the following, we present the observed themes.

\paragraph{Theme 1: Adaptive Learning and Personalized Skill Development.} 
The theme of adaptive and personalized learning emerges prominently from the data, highlighting a growing shift towards learner-centric education that leverages technology to enhance skill acquisition, engagement, and individualized support. Participants advocated for tools that dynamically adapt content based on learners' prior knowledge, enabling real-time assessment and feedback. In parallel, there is an inclination toward AI-driven tools that can personalize learning pathways, guiding learners through tailored courses, recommending optimal resources and activities, and monitoring learning over time [\textit{"I (educator) want an AI tool that continuously monitors the progress of students, and provides insights into their gained competences or improvements"}].

\paragraph{Theme 2: Monitoring and Enhancing Learner Motivation and Engagement.}
This theme of monitoring and fostering learner motivation and engagement highlights participants’ strong interest in sustaining and enhancing students' enthusiasm. They envision AI solutions that not only monitor engagement but also assist educators in creating captivating, relevant, and personalized learning experiences that stimulate and sustain learners' interest [\textit{"The challenges are to create content that will animate the majority"}]. Participants emphasized AI's potential to create such environments, which ultimately contributes to more effective educational outcomes [\textit{"Some kind of short check..to be positive and encouraging so that students continue their progress"}].

\paragraph{Theme 3: Enhancing Pedagogy via Data-driven Design.}
Amid the growing interest in AI integration, the data from the workshop revealed a strong emphasis on enhancing pedagogy via data-informed course design. Participants articulated a vision for AI tools that not only support the delivery of content but also provide meaningful pedagogical guidance, such as in forming effective collaborative groups [\textit{"..recommend meaningful collaboration groups of students based on aggregated quality characteristics and their learning patterns"}]. In tandem, they emphasized the importance of aligning pedagogy with learners’ evolving needs, tracking progress in real time, and ensuring that materials remain relevant [\textit{"I need the tool to collects and analyze learners' expectations for the course...to adapt and customize the course content to better meet their needs and objectives"}]. 

\paragraph{Theme 4: Enhancing Feedback in Collaborative Inquiry-Based Learning.}
This theme expresses the participants' interest in AI solutions that support group work by analyzing contributions, offering suggestions, and providing diverse perspectives to enrich the learning process. For instance, participants highlighted the desire for tools that provides feedback during inquiry-based activities [\textit{"I would like to have a solution that helps my students when working on inquiry-based approaches and in groups"}], as well as offer reflective insights based on engagement data [\textit{"I (teacher) want to receive feedback based on data during lessons, so that I can improve engagement and learning experience"}]. 

\section{Discussion}
This study employs a dual thematic analysis approach to provide a theoretical framework to analyse how user needs align or clash with their expectations. The deductive analysis outlined participants' needs from an activity system perspective \cite{Engeström_2014}, revealing issues like educators’ resistance to change, lack of institutional support, and desire for better tools and training. In tandem, the inductive analysis uncovered their emergent expectations, many of which resonated strongly with identified needs.

However, our cross-examination also unveiled contradictions, i.e, historically accumulating structural tensions within an activity system \cite{Engeström2001}. Users expect rich data-driven personalization from AI, yet simultaneously voice concerns about data privacy, transparency, and the credibility of AI-driven assessments, a sentiment also echoed by Chounta et. al., \cite{Chounta_2024}. Similarly, educators desire AI to offload routine and even high-level tasks (e.g., automating feedback), but lack trust in AI’s pedagogical judgment. 

Teachers are expected to integrate AI and 21st-century skills in their teaching, but no institutional support is provided. This tension signals that, without clarifying new responsibilities (who trains the AI, evaluates its recommendations), simply introducing AI could exacerbate existing role ambiguities. Educators were generally enthusiastic about emerging technologies such as mixed-reality systems for their potential to support authentic practice, but were reluctant due to the mixed evidence regarding their efficacy \cite{Lee_2023_Role} and the complexity in their implementation \cite{Limbu_Holo_2023}. Similarly, teachers tend to agree on the importance of facilitating 21st-century skills development but remark on the lack of proper pedagogy.

By synthesizing these findings, we derive several fundamental suggestions to address the above intersections and conflicts. First, these systems should embed support for various external factors, such as ethical safeguards, to reduce adoption effort. These systems should adopt privacy by design to earn \& maintain trust and ensure transparency. Second, hybrid intelligence approaches should be adopted to clarify the roles of AI and teachers \cite{Thomas_2024_teacher}. AI can handle routine analytics (monitoring engagement) and offer pedagogical recommendations without undermining teachers' agency. Teachers generally perceive AI in education positively, but have limited knowledge about AI and therefore must be supported by the institution \cite{chounta_exploring_2022}. In summary, these requirements underscore the importance of teachers' perception of AI in education, which heavily influences its sustained adoption. By addressing these contradictions – essentially designing for the “expansion” of current practice rather than disruption – AI tools can align with educators’ workflows and lead to wide adoption.

\section{Conclusion}

We presented the participatory approach used to elicit user requirements for AI-enhanced learning environments. Through a two-phase workshop, we captured and analysed the needs and expectations of stakeholders, while also unveiling the contradictions and challenges that shape the integration of AI in education. By cross-examining their needs and expectations, we extracted user requirements and suggestions to foster a broad adoption of AI solutions. Our findings emphasize the critical role of human-centered, activity and context-aware design in ensuring that AI technologies meaningfully support, rather than disrupt, educational practices.

\begin{credits}
\subsubsection{\ackname} \label{secAckowledge} This research was developed within the augMENTOR (“Augmented Intelligence for Pedagogically Sustained Training and Education”) project, co-funded by the European Commission under HORIZON-CL2-2021-TRANSFORMATIONS-01-05, project number: 101061509.

\subsubsection{\discintname}
The authors have no competing interests to declare that are relevant to the content of this article.
\end{credits}
%
%
%
\bibliographystyle{splncs04}
\bibliography{Ref}

\begin{thebibliography}{10}
\providecommand{\url}[1]{\texttt{#1}}
\providecommand{\urlprefix}{URL }
\providecommand{\doi}[1]{https://doi.org/#1}

\bibitem{braun2021thematic}
Braun, V., Clarke, V.: Thematic Analysis: A Practical Guide. SAGE Publications (2021)

\bibitem{braun_using_2006}
Braun, V., Clarke, V.: Using thematic analysis in psychology. Qualitative Research in Psychology  \textbf{3}(2),  77--101 (Jan 2006). \doi{10.1191/1478088706qp063oa}

\bibitem{chounta_exploring_2022}
Chounta, I.A., Bardone, E., Raudsep, A., Pedaste, M.: Exploring {Teachers}’ {Perceptions} of {Artificial} {Intelligence} as a {Tool} to {Support} their {Practice} in {Estonian} {K}-12 {Education}. International Journal of Artificial Intelligence in Education  \textbf{32}(3) (Sep 2022). \doi{10.1007/s40593-021-00243-5}

\bibitem{Chounta_2024}
Chounta, I.A., Limbu, B., van~der Heyden, L.: Exploring the Methodological Contexts and Constraints of Research in Artificial Intelligence in Education, p. 162–173. Springer Nature Switzerland (2024). \doi{10.1007/978-3-031-63028-6_13}

\bibitem{clemmensen_making_2016}
Clemmensen, T., ~, Victor, K., , Nardi, B.: Making {HCI} theory work: an analysis of the use of activity theory in {HCI} research. Behaviour \& Information Technology  \textbf{35}(8),  608--627 (Aug 2016). \doi{10.1080/0144929X.2016.1175507}, publisher: Taylor \& Francis

\bibitem{dickler2022interdisciplinary}
Dickler, R., Dudy, S., Mawasi, A., Whitehill, J., Benson, A., Corbitt, A.: Interdisciplinary approaches to getting ai experts and education stakeholders talking. In: International Conference on Artificial Intelligence in Education. pp. 115--118. Springer (2022)

\bibitem{ElHelou_Tzagarakis_Gillet_Karacap_Yu_2008}
El~Helou, S., Tzagarakis, M., Gillet, D., Karacapilidis, N., Yu, C.M.: Participatory design for awareness features: Enhancing interaction in communities of practice. Proceedings of the International Conference on Networked Learning  \textbf{6},  523–530 (May 2008). \doi{10.54337/nlc.v6.9370}

\bibitem{Engeström2001}
Engeström, Y.: Expansive learning at work: Toward an activity theoretical reconceptualization. Journal of Education and Work  \textbf{14}(1),  133--156 (2001). \doi{10.1080/13639080020028747}

\bibitem{Engeström_2014}
Engeström, Y.: Learning by Expanding: An Activity-Theoretical Approach to Developmental Research. Cambridge University Press, 2 edn. (2014)

\bibitem{Thomas_2024_teacher}
Fr{{\o}}sig, T.B., Romero, M.: {Teacher agency in the age of generative AI: towards a framework of hybrid intelligence for learning design}. In: {IRMBAM 2024}. {IPAG}, Nice, France (Jul 2024), \url{https://hal.science/hal-04639071}

\bibitem{holmes2022artificial}
Holmes, W., Persson, J., Chounta, I.A., Wasson, B., Dimitrova, V.: Artificial intelligence and education: A critical view through the lens of human rights, democracy and the rule of law. Council of Europe (2022)

\bibitem{kensing1998participatory}
Kensing, F., Blomberg, J.: Participatory design: Issues and concerns. Computer supported cooperative work (CSCW)  \textbf{7},  167--185 (1998)

\bibitem{Lavidas_2024_determinants}
Lavidas, K., Voulgari, I., Papadakis, S., Athanassopoulos, S., Anastasiou, A., Filippidi, A., Komis, V., Karacapilidis, N.: Determinants of humanities and social sciences students’ intentions to use artificial intelligence applications for academic purposes. Information  \textbf{15}(6) (2024). \doi{10.3390/info15060314}

\bibitem{Lee_2023_Role}
Lee, Y., Limbu, B., Rusak, Z., Specht, M.: Role of {Multimodal} {Learning} {Systems} in {Technology}-{Enhanced} {Learning} ({TEL}): {A} {Scoping} {Review}. In: Viberg, O., Jivet, I., Muñoz-Merino, P., Perifanou, M., Papathoma, T. (eds.) Responsive and {Sustainable} {Educational} {Futures}. pp. 164--182. Springer Nature Switzerland, Cham (2023). \doi{10.1007/978-3-031-42682-7_12}

\bibitem{Limbu_Holo_2023}
Limbu, B., van Roijen, R., Beerens, M., Specht, M.: {HoloLearn}: {Towards} a {Hologram} {Mediated} {Hybrid} {Education}. In: Dascalu, M., Mealha, O., Virkus, S. (eds.) Smart {Learning} {Ecosystems} as {Engines} of the {Green} and {Digital} {Transition}. pp. 117--132. Springer Nature, Singapore (2023). \doi{10.1007/978-981-99-5540-4\_7}

\bibitem{muller_participatory_2012}
Muller, M.J., Druin, A.: Participatory {Design}: {The} {Third} {Space} in {Human}–{Computer} {Interaction}. In: Human {Computer} {Interaction} {Handbook}. CRC Press, 3 edn. (2012), num Pages: 29

\bibitem{spinuzzi2005methodology}
Spinuzzi, C.: The methodology of participatory design. Technical communication  \textbf{52}(2),  163--174 (2005)

\bibitem{topali2025designing}
Topali, P., Ortega-Arranz, A., Rodr{\'\i}guez-Triana, M.J., Er, E., Khalil, M., Ak{\c{c}}ap{\i}nar, G.: Designing human-centered learning analytics and artificial intelligence in education solutions: a systematic literature review. Behaviour \& Information Technology  \textbf{44}(5),  1071--1098 (2025)

\bibitem{uden_activity_2024}
Uden, L., Ching, G.: Activity {Theory}-based {Ecosystem} for {Artificial} {Intelligence} in {Education} ({AIED}). International Journal of Research Studies in Education  \textbf{13},  41--54 (May 2024). \doi{10.5861/ijrse.2024.24000}

\end{thebibliography}

\end{document}